\documentclass[10pt,twocolumn,letterpaper]{article}

\usepackage{cvpr} 
\usepackage{times}
\usepackage{epsfig}
\usepackage{graphicx}
\usepackage{amsmath}
\usepackage{amssymb}
\usepackage[numbers,sort&compress]{natbib}
\usepackage{tabularray}
\usepackage{adjustbox}
\usepackage{doi}
\usepackage{multirow}

\begin{document}

\title{Privatization of Synthetic Gaze: Attenuating State Signatures in Diffusion-Generated Eye Movements}

\author{
Kamrul Hasan, Oleg V. Komogortsev\\
Texas State University, San Marcos, Texas, USA\\
{\tt\small \{kamrul.hasan, ok\}@txstate.edu}
}

\maketitle
\thispagestyle{empty}

\begin{abstract}
The recent success of deep learning (DL) has enabled the generation of high-quality synthetic data, advancing the development of data-driven biometric applications. Among various biometric modalities, eye movement sequences have emerged as a promising behavioral biometric. However, gaze data also raises privacy concerns because it may encode individuals’ internal states, such as fatigue, emotional load, and stress. Ideally, synthetic gaze data should preserve the signal quality of real recordings, including identity features, while removing or attenuating privacy-sensitive, state-related attributes to reduce risks of personal state exposure. Many recent DL-based generative models focus on replicating real gaze trajectories but do not explicitly evaluate whether generated signals retain subjective-state information. In this work, we examine a recent diffusion-based gaze synthesis approach by analyzing the correlations between synthetic gaze features and subjective reports, including fatigue and other self-reported states. Our results show that these correlations are weaker and less stable in synthetic gaze than in real gaze, suggesting the attenuation of state-related signatures under the evaluated protocol. At the same time, synthetic gaze preserves essential signal characteristics similar to real data, supporting its potential use in privacy-aware gaze-based applications.
\end{abstract}


\section{Introduction}
Eye-tracking is a widely used biometric technology for studying human cognition and behavior \cite{hayhoe2005eye, liu2021effectiveness} and has broad applications in clinical assessment \cite{haller2022eye, thanarajan2023eye}, education \cite{sun2017application}, user authentication \cite{lohr2022ekyt, lohr2025gaze}, and augmented reality or virtual reality (AR/VR) interfaces \cite{jin2024eye}. Unlike other biometric modalities, such as face \cite{zhao2003face}, fingerprint \cite{maltoni2009handbook}, and gait \cite{uddin2024horizontal}, eye-tracking offers advantages by being contactless and can provide high recognition accuracy, resistance to spoofing attempts, and temporal consistency without requiring physical touch or full facial exposure. Moreover, high-frequency eye movement recordings capture rich oculomotor dynamics and detailed eye movement patterns \cite{griffith2021gazebase}. Despite these advantages, real eye-tracking recordings are expensive to collect, difficult to share, and raise substantial privacy concerns because they can reveal both biometric identity and privacy-sensitive internal states such as fatigue, stress, and emotional load \cite{steil2019privacy}. To address data scarcity and sharing constraints, synthetic gaze data has emerged as a scalable alternative to real eye-tracking recordings \cite{prasse2023sp}.

Recent works \cite{prasse2023sp, jiao2024diffeyesyn} have shown that generative models, including diffusion \cite{ho2020denoising} and generative adversarial networks (GANs) \cite{goodfellow2020generative}, can synthesize gaze sequences that closely resemble real recordings. These methods also appear effective at preserving individual-specific patterns when evaluated using standard measures such as spatial accuracy and precision \cite{aziz2024evaluation, hasan2025quantitative}. However, most prior work focuses on signal realism and biometric utility, with limited attention to whether synthetic gaze also retains internal-state signatures. This issue is important because high-fidelity synthetic gaze may preserve temporal dynamics, event structure, inter-individual variability, and potentially state-dependent variability that are relevant to downstream tasks \cite{qian2025we}. Therefore, signal fidelity and privacy awareness may be in tension: a model that more accurately reproduces real gaze dynamics may also retain features associated with fatigue, cognitive load, or emotional state.
In this work, we examine whether diffusion-generated synthetic gaze preserves the associations between eye-movement features and subjective self-reports. 
We employed an updated DiffEyeSyn framework \cite{hasan2025quantitative} conditioned on a 25 Hz identity-removed velocity signal and compact 128-D embeddings \cite{lohr2022ekyt}.
After generating subject-specific synthetic gaze sequences, we extract interpretable eye movement features, including saccade rate and fixation drift, and compute their correlations with subjective ratings of eye tiredness, mental tiredness, and perceived task difficulty. This analysis allows us to evaluate whether synthetic gaze attenuates measured state-related associations while preserving task-relevant oculomotor structure. To the best of our knowledge, this is the first study to analyze correlations between diffusion-generated synthetic gaze features and subjective self-reports. The major contributions of our study are summarized as follows:

\begin{enumerate}

\item We employ an updated diffusion-based gaze synthesis framework to generate subject-specific synthetic gaze signals and analyze them using an objective eye-movement feature-extraction pipeline.
\item We introduce a correlation-based evaluation protocol to assess whether synthetic gaze retains subjective-state signatures through associations between objective eye movement features and subjective self-reports.

\item We provide empirical insights for designing privacy-aware synthetic gaze data by showing that diffusion-generated signals can preserve task-relevant oculomotor structure while attenuating state-related associations.


\end{enumerate}

\section{Related Work}
\subsection{Synthetic Gaze Generation}
Synthetic gaze generation approaches can be broadly classified into statistical methods and deep learning-based methods. Earlier studies \cite{lee2002eyes, duchowski2015modeling, duchowski2016eye, yeo2012eyecatch} primarily relied on training-free statistical simulators that generated either average scanpath animations or photorealistic eye movement sequences based on simplified oculomotor dynamics. 
Although such approaches are easy to implement and do not require training, they typically produce smoothed or overly averaged gaze signals or images and often fail to capture fine-grained oculomotor characteristics.

In contrast, deep learning-based approaches \cite{simon2016automatic, assens2018pathgan, goodfellow2020generative, jiao2023supreyes, hasan2026diffusion} mostly employ generative models, such as generative adversarial networks (GANs), variational autoencoders (VAEs), and diffusion models to synthesize gaze data. These methods can better capture key signal characteristics, including microsaccades, fixations, and other fine-grained oculomotor events. 
For example, Simon et al. \cite{simon2016automatic} used a CNN-LSTM model to generate scanpaths from static images, while SP-EyeGAN \cite{prasse2023sp} generated fixations and saccades separately and then combined them to construct complete gaze signals.
More recently, diffusion-based approaches such as DiffGaze \cite{jiao2025diffgaze} and DiffEyeSyn \cite{jiao2024diffeyesyn} have been proposed to produce more realistic gaze signals. Building on this line of work, we utilize an updated version of DiffEyeSyn \cite{hasan2025quantitative} as the generative backbone for subject-specific synthetic gaze analysis.

\subsection{Privacy Preservation in Gaze Data}
Eye-tracking data are privacy-sensitive because fine-grained gaze trajectories can encode both biometric identity and cognitive or affective state information \cite{aziz2023assessing}. Prior work has shown that eye movements can support user authentication and may also reveal fatigue, attention, and mental health-related traits \cite{kroger2020does}, raising legal and ethical concerns under regulations such as the GDPR \cite{david2022your}. To reduce such risks, several signal-level privacy techniques have been studied, including differential privacy for eye-movement features \cite{bozkir2021differential}, randomized perturbations \cite{liu2019differential}, filter-based approaches \cite{raju2025real}, and noise-injection frameworks \cite{aziz2025privacy}. 

Within this broader context, synthetic eye movement data provide a promising direction for reducing dependence on real gaze recordings while preserving useful signal characteristics \cite{duchowski2016eye, jiao2024diffeyesyn}. Recent work by Qian et al. \cite{qian2025we} argues that synthetic oculomotor data should be evaluated not only for signal realism but also for temporal dynamics, event structure, inter-individual variability, and state-dependent variability. However, most existing synthetic gaze studies do not empirically test whether generated signals preserve associations with subjective internal states such as fatigue, illness, eye tiredness, or mental stress. We address this gap by evaluating whether diffusion-generated gaze attenuates feature-level associations with subjective reports while retaining essential task-level signal structure.


\begin{figure*}[ht]
    \centering
\includegraphics[width=\linewidth]{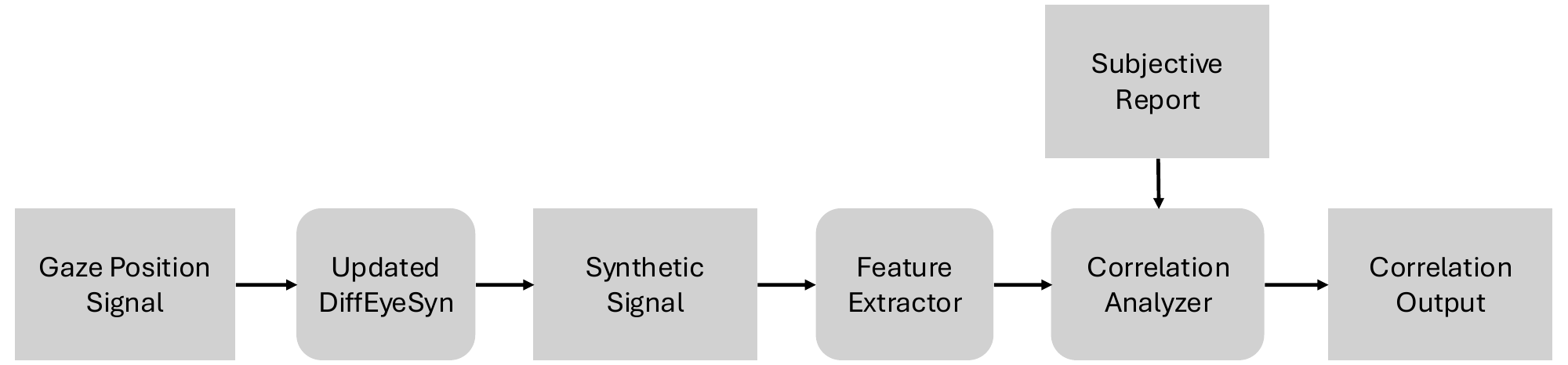}
    \caption{Overview of the proposed architecture for analyzing the correlation between subjective reports and objective eye movement features.
    }
    \label{fig:main_architecture}
\end{figure*}

\section{Proposed Method}
Figure~\ref{fig:main_architecture} summarizes the framework. Updated DiffEyeSyn first generates subject-specific synthetic gaze from real positional input. The feature extractor then computes objective oculomotor descriptors, and the correlation analyzer quantifies their associations with subjective reports.

\subsection{Synthetic Gaze Generation}
We employed an updated DiffEyeSyn architecture \cite{hasan2025quantitative}, which is a conditional denoising diffusion probabilistic model (DDPM) \cite{ho2020denoising, zhang2023adding}, to synthesize subject-specific eye-movement sequences from identity-removed velocity signals and compact user embeddings. Given a real positional gaze signal ($p$), a preprocessing module first converts it into a velocity signal ($v$) and an identity-removed velocity signal ($v_0$). In the updated setting, $v_0$ is obtained by downsampling the original signal to 25 Hz and then upsampling it back to 1000 Hz. This choice is motivated by Percentage of Variance Accounted For (PVAF) analysis \cite{raju2021determining}, which indicates that the 0--25 Hz band preserves most saccadic signal variance while attenuating higher-frequency components associated with identity-specific information. Thus, $v_0$ is used to preserve the overall gaze structure while reducing identity- and state-related cues.


During diffusion, Gaussian noise is added to $v$ at timestep $t$ to produce a noisy sample $x_t$. Subject-specific conditioning is provided by a pre-trained EKYT encoder \cite{lohr2022ekyt}, which extracts a compact 128-dimensional embedding ($z$) from $v$ with a single fold model. Compared with the original 512-dimensional setting, the reduced embedding dimension has less impact on the generator. The conditioning pair $(v_0,z)$ is injected into the denoiser $\varepsilon_\theta(x_t,t,\mathrm{cond})$ using Feature-wise Linear Modulation (FiLM), which adaptively scales and shifts intermediate feature representations during denoising. Later, the denoiser predicts the noise term ($\hat{\varepsilon}_t$), which is used to recover the denoised velocity signal $\hat{v}$ and define the reverse transition $p_\theta(x_{t-1}\mid x_t,\mathrm{cond})$. Overall, the model is trained with a weighted objective combining noise-prediction loss and identity-guidance loss, where the latter aligns the embeddings of $\hat{v}$ and $v$ using cosine similarity.

\subsection{Correlation Analysis with Subjective Reports}
To assess how much subjective state information is preserved in synthetic gaze, we adopt a correlation-based baseline \cite{qian2025we}. In this framework, subjective reports are treated as behavioral readouts of participants’ internal states, such as fatigue, mental load, and perceived difficulty. For each synthetic gaze sequence, eye movement events are first segmented using the algorithm described in \cite{rigas2018study}. We then extract objective eye movement features using the event-identification and feature-extraction methods proposed in \cite{friedman_novel_2018, friedman_eye_2022}. These features are defined either as the rate of a given event type (e.g., saccades per second) or as the median value of multi-valued measures.

As illustrated in Figure \ref{fig:main_architecture}, once a synthetic gaze sequence is generated, the feature extraction module derives an interpretable set of eye movement features, including saccade rate (number of rapid eye movements between fixation points), frequency, amplitude (magnitude of saccade movements), fixation rate (number of moments where the gaze is held steady), drift (slow movement of the eye during fixation), and measures related to the stability of eye position. From this set, we select 58 features representing objective eye movement characteristics, following the definitions used in the baseline study \cite{qian2025we}. The correlation analyzer then utilizes these extracted features and subjective reports (i.e., participants’ self-reported experiences) to quantify the relationship between synthetic gaze and subjective state for each task and subject. As the subjective ratings are non-normally distributed and exhibit floor effects, we compute Spearman rank correlations \cite{zar2005spearman} between each subjective measure and each eye movement feature. 
In this study, correlation strength is used as an empirical proxy for internal-state leakage. Therefore, reduced feature-rating correlations indicate state-signature attenuation under the evaluated feature set and statistical protocol, not anonymization, differential privacy, or a formal guarantee against state-inference attacks.

\section{Experiments}
\subsection{Dataset}
We utilized the publicly available GazeBase \cite{griffith2021gazebase} dataset, collected with an EyeLink 1000 eye tracker at 1000 Hz using left-eye recordings. GazeBase contains nine rounds, each with two sessions, where participants performed seven tasks: fixation (FXS), horizontal saccades (HSS), random oblique saccades (RAN), reading (TEX), free viewing of cinematic videos (VD1 and VD2), and gaze-driven gaming (BLG). For training the generative model, we used recordings from 263 participants across all rounds, yielding 63,161 non-overlapping 5-second gaze sequences. Signal-quality evaluation was performed on a disjoint, held-out set of 59 participants from round 6. At test time, the model generated non-overlapping 5-second synthetic gaze windows, which were concatenated for each subject, session, and task to reconstruct complete synthetic gaze signals with the same temporal length as the corresponding real recordings.

Consistent with the baseline subjective-state analysis \cite{qian2025we}, we focus on HSS, RAN, and TEX because these tasks align with the baseline protocol and cover complementary oculomotor behaviors: controlled horizontal saccades, randomized oblique saccades, and reading. For the correlation analysis, we generated synthetic gaze data for the same rounds, sessions, tasks, and subjects used in the baseline subjective-report subset. We also incorporated the subjective ratings collected in GazeBase \cite{griffith2021gazebase}, where participants reported overall task difficulty (OverDiff), mental tiredness (Mentally), and eye tiredness (TiredEyes) on a 1--7 Likert scale, with 1 denoting ``not at all tired or difficult'' and 7 denoting ``extremely tired or difficult.'' This protocol ensures direct comparability with the baseline analysis.



\subsection{Implementation Details}
We trained updated DiffEyeSyn on 5-second gaze windows sampled at 1,000 Hz (5,000 samples per window), using a diffusion process with $T=50$ discrete time steps and a linear noise schedule ranging from 0.0001 to 0.05 for both the forward and reverse processes. The AdamW optimizer \cite{loshchilov2017decoupled} was used with a learning rate of 0.0002 and a batch size of 32, and the model was trained for 450 epochs. All experiments were conducted in Python 3.9.12 with CUDA 12.4, PyTorch 2.1.0, PyTorch Lightning 1.9.5, and TensorFlow 2.14.0 on an NVIDIA RTX A6000 GPU (48 GB GDDR6). For the correlation analysis, we utilized MATLAB Version 25.2.0.3055257 (R2025b) with the necessary toolboxes installed. The source code is available on the Texas State Digital Collections Repository at \url{https://hdl.handle.net/10877/25138}.


\begin{figure*}[ht]
    \centering
    \includegraphics[width=\linewidth]{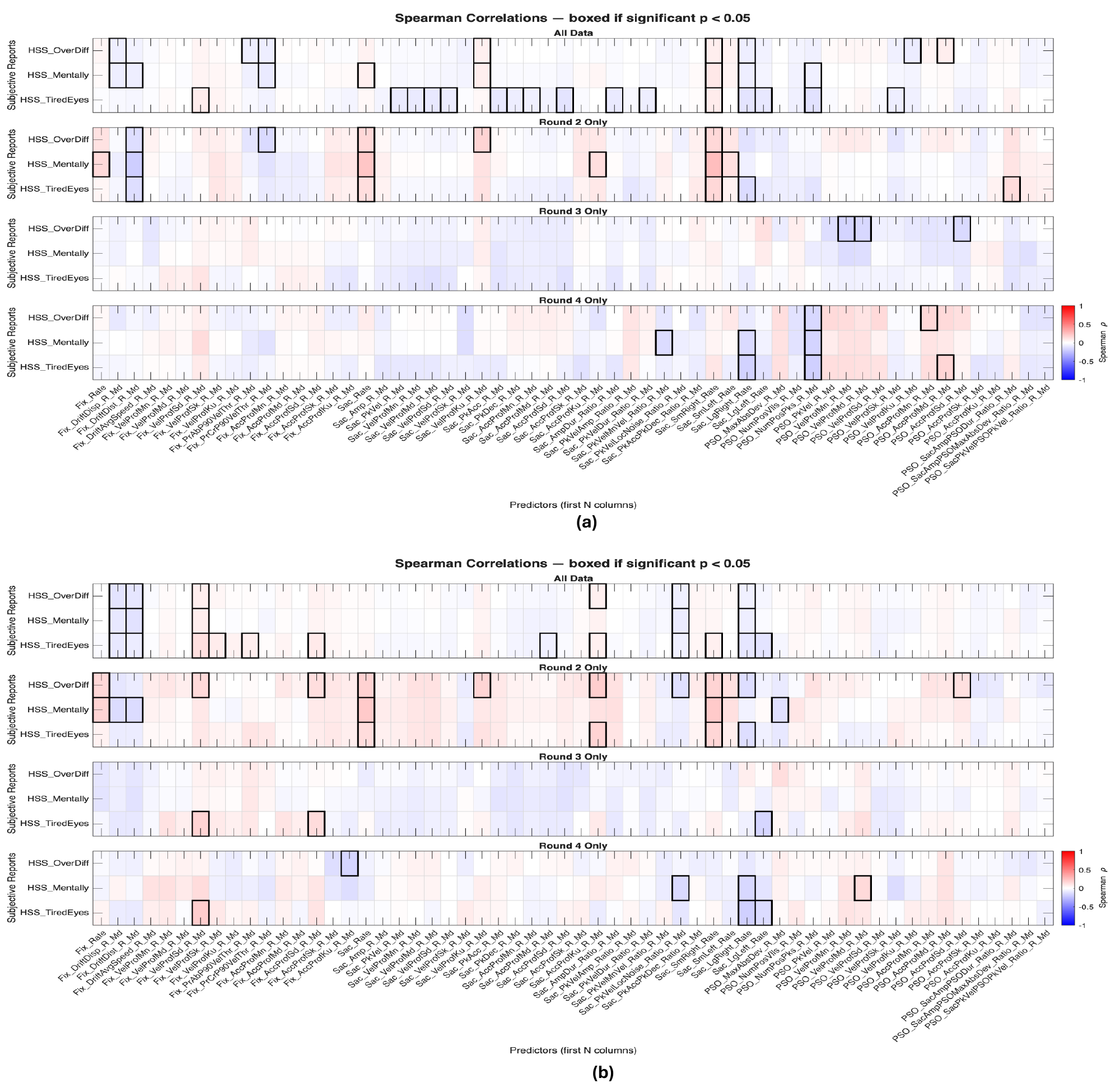}
    \caption{Spearman correlations between gaze-derived features and subjective ratings for the HSS task: (a) real gaze, (b) synthetic gaze.}
    \label{fig:hss}
\end{figure*}

\begin{figure*}[ht]
    \centering
    \includegraphics[width=\linewidth]{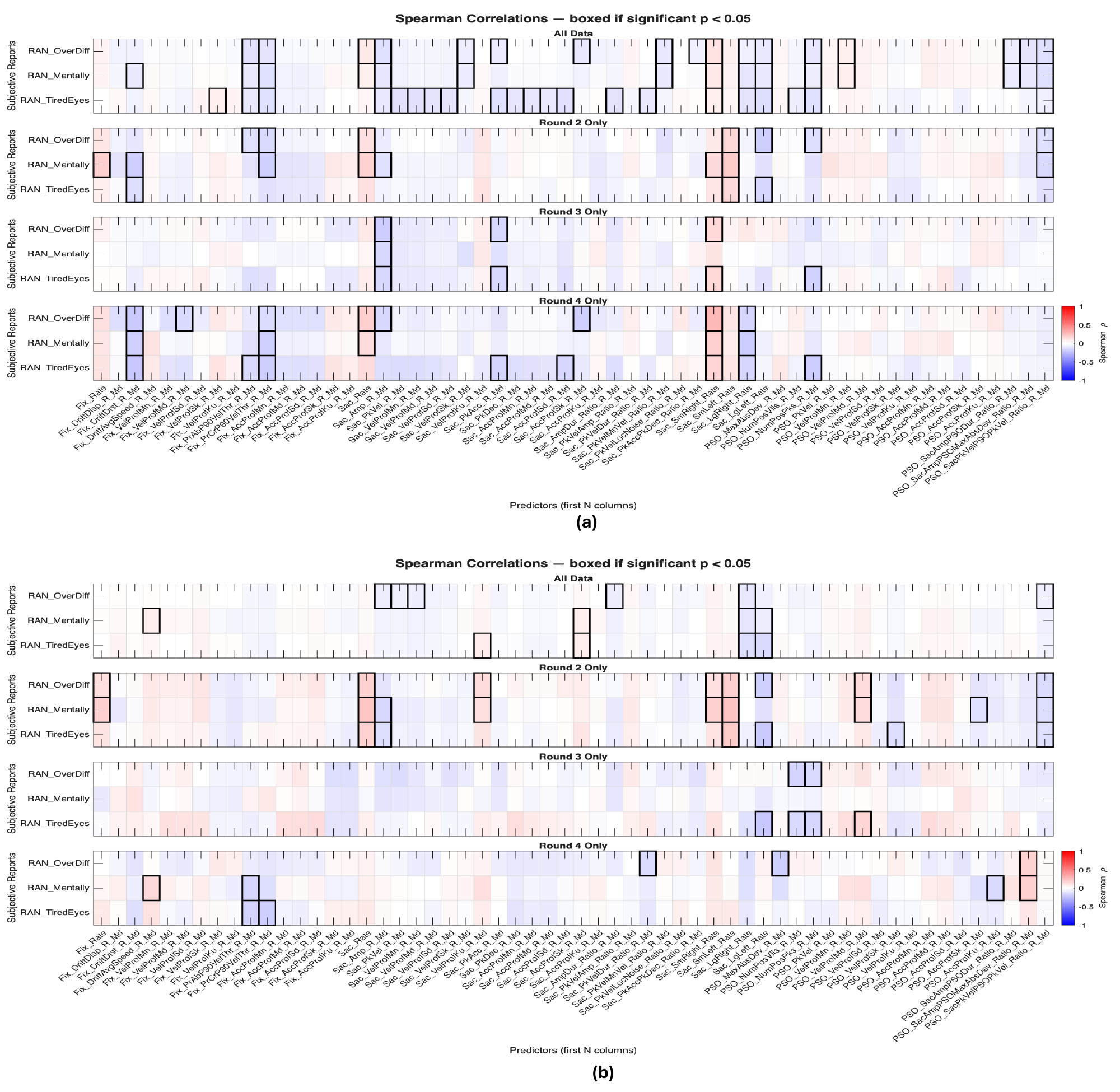}
    \caption{Spearman correlations between gaze-derived features and subjective ratings for the RAN task: (a) real gaze, (b) synthetic gaze.}
    \label{fig:ran}
\end{figure*}

\begin{figure*}[ht]
    \centering
    \includegraphics[width=\linewidth]{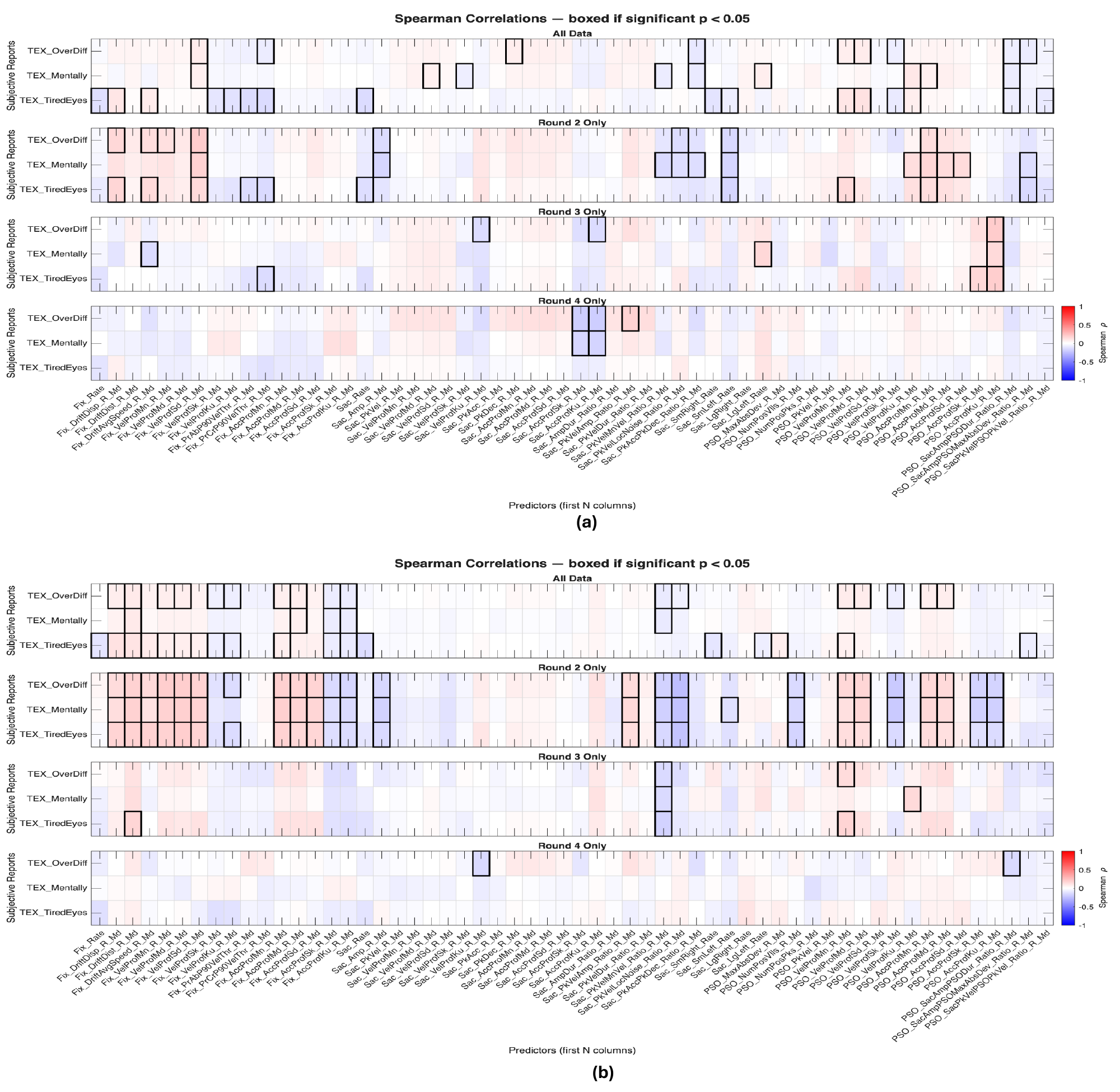}
    \caption{Spearman correlations between gaze-derived features and subjective ratings for the TEX task: (a) real gaze, (b) synthetic gaze.}
    \label{fig:tex}
\end{figure*}

\section{Results}

\subsection{Synthetic Signal Quality}
After training, updated DiffEyeSyn achieved spatial accuracy errors of approximately 3.73 and 4.06 degrees of visual angle (DVA) on HSS and RAN, respectively. Although these errors may be limiting for applications requiring precise point-of-gaze localization, they are acceptable for our goal of evaluating privacy-aware synthetic gaze at the signal and biometric-feature level. DiffEyeSyn also achieved root-mean-square (RMS) precision of 0.01 for both HSS and RAN, matching the stability of the ground-truth signals and indicating minimal fixation jitter. The TEX task was excluded from the spatial accuracy and precision analysis because it does not have a ground-truth stimulus trajectory in the dataset.

To further assess signal fidelity, we extracted embeddings from synthetic and real gaze sequences using the pre-trained EKYT encoder \cite{lohr2022ekyt} and computed cosine similarity in the embedding space. As shown in Table \ref{tab:cosine_similarity}, the cosine similarities are $0.956 \pm 0.015$ for HSS, $0.956 \pm 0.016$ for RAN, and $0.951 \pm 0.021$ for TEX. Together, the spatial accuracy, spatial precision, and embedding similarity results indicate that DiffEyeSyn preserves stable task-level trajectories and subject-relevant oculomotor structure across both saccadic and reading behaviors.

\begin{table}[h]
\centering
\caption{Cosine similarities between the EKYT embeddings of real and synthetic eye movement data.}
\label{tab:cosine_similarity}
\begin{adjustbox}{width=0.35\textwidth}
{\small
\renewcommand{\arraystretch}{1}
\begin{tabular}{ccc} 
\hline
\multicolumn{3}{c}{\textbf{Task}}              \\ 
\hline
\textbf{HSS}  & \textbf{RAN}  & \textbf{TEX}   \\ 
\hline
0.956 ± 0.015 & 0.956 ± 0.016 & 0.951 ± 0.021  \\
\hline
\end{tabular}
}
\end{adjustbox}
\end{table}

\subsection{Correlation between Eye Movement Features and Subjective Reports}
This section presents correlation analysis of synthetic gaze sequences across three tasks: HSS, RAN, and TEX. For each task, we compute Spearman rank correlations separately for (i) all rounds pooled and (ii) rounds 2-4 individually. In the correlation heatmaps, boxed cells indicate coefficients with uncorrected $p < 0.05$. We compute correlations using all 58 gaze-derived features, following baseline \cite{qian2025we}. These features include fixation (\textit{Fix\_*}), saccade (\textit{Sac\_*}), and post-saccadic oscillation (\textit{PSO\_*}) descriptors; \textit{H}, \textit{V}, and \textit{R} denote horizontal, vertical, and resultant components, respectively, where $R = \sqrt{H^2 + V^2}$, and suffix \textit{Md} denotes the median.


For the HSS task (Figure~\ref{fig:hss}), synthetic data exhibit weaker and sparser associations between subjective reports and gaze-derived features than real data. While several significant correlations remain in the synthetic heatmaps, they are fewer and less consistent across rounds than those observed in the real data for any of the three ratings (i.e., OverDiff, Mentally, and TiredEyes). Some isolated fixation-related associations appear in individual rounds (e.g., median fixation velocity profile std dev (Fix\_VelProfSd\_R\_Md)) in rounds 3 and 4 for TiredEyes, but these effects do not form a stable pattern in the pooled analysis. Overall, synthetic HSS sequences only partially preserve the relationship between gaze features and subjective ratings.

A similar pattern is observed for the RAN task (Figure~\ref{fig:ran}). In synthetic data, significant correlations are fragmented and mostly limited to isolated features in individual rounds, with little consistency across rounds or in the pooled analysis. For example, the median saccade amplitude resultant (i.e., Sac\_Amp\_R\_Md) is consistent across all rounds in real data, whereas in synthetic data, the correlation is absent. Overall, compared with real data, the broader correlation structure is notably weaker, indicating that the associations between gaze-derived features and subjective ratings are weakly preserved in the synthetic RAN sequences.


Among the three tasks, TEX (Figure~\ref{fig:tex}) exhibits the richest correlation structure in the synthetic data. In the pooled analysis, several features remain significantly associated with perceived difficulty (OverDiff) and tired eyes (TiredEyes). Round 2 also shows a localized cluster of significant effects. However, these patterns weaken substantially in rounds 3 and 4. In these rounds, only a few features remain significant. For example, the resultant kurtosis of the median saccade acceleration profile (Sac\_AccProfKu\_R\_Md) is completely absent in the synthetic correlation graph. Taken together, these results suggest that synthetic data retain some task-dependent structure, especially for TEX. Overall, though, they show weaker, less stable feature-rating associations than the real data across all three tasks. This indicates that the diffusion-based generative model preserves overall oculomotor characteristics but substantially diminishes the systematic link between eye-movement features and subjective states.

In summary, the results exhibit four main patterns: (a) high similarity between real and synthetic embeddings, indicating preserved task-relevant oculomotor structure; (b) sparse and inconsistent synthetic feature-rating associations for HSS and RAN; (c) stronger residual task-dependent structure for TEX, especially in pooled analysis and round 2; and (d) overall weaker and less reliable synthetic associations with subjective reports, indicating attenuation rather than complete removal of state-related signatures.

\subsection{State-Inference Validation}
To complement the correlation analysis, we performed a binary state-inference evaluation using the same 58 gaze-derived features. For each subjective report, namely OverDiff, Mentally, and TiredEyes, ratings were binarized into low and high groups, where scores of 1--3 were treated as low and scores of 4--7 were treated as high. We trained a class-balanced logistic regression classifier and evaluated it using subject-disjoint folds to avoid identity leakage. Balanced accuracy was used as the main metric because the subjective labels were imbalanced.

As shown in Table~\ref{tab:state_inference}, performance remains close to chance for both real and synthetic gaze across most tasks and states. HSS and RAN show only small real-synthetic differences. TEX shows lower synthetic recoverability for Mentally but higher synthetic recoverability for OverDiff, indicating that state recoverability is weak and inconsistent under this binary classification protocol. Therefore, the classifier-based validation should be interpreted as an attacker-oriented sanity assessment that complements the correlation analysis, rather than as direct proof of state-signature removal or formal privacy protection.


\begin{table}[h]
\centering
\caption{Binary state-inference performance with subject-disjoint logistic regression. Balanced accuracy is reported; $\Delta>0$ means higher recoverability from real gaze.}
\label{tab:state_inference}
\begin{adjustbox}{width=0.35\textwidth}
{\small
\renewcommand{\arraystretch}{1}
\begin{tabular}{llccc}
\hline
Task & State & Real & Synthetic & $\Delta$ \\
\hline
\multirow{3}{*}{HSS} 
& Mentally  & 0.465 & 0.423 &  0.041 \\
& OverDiff  & 0.498 & 0.490 &  0.009 \\
& TiredEyes & 0.515 & 0.543 & -0.028 \\
\hline
\multirow{3}{*}{RAN} 
& Mentally  & 0.460 & 0.464 & -0.004 \\
& OverDiff  & 0.497 & 0.468 &  0.029 \\
& TiredEyes & 0.512 & 0.514 & -0.002 \\
\hline
\multirow{3}{*}{TEX} 
& Mentally  & 0.546 & 0.469 &  0.077 \\
& OverDiff  & 0.444 & 0.555 & -0.111 \\
& TiredEyes & 0.517 & 0.501 &  0.016 \\
\hline
\end{tabular}
}
\end{adjustbox}
\end{table}


\subsection{Ablation Analysis}
To assess the effect of the updated conditioning configuration, we compare the original setting using a 20 Hz identity-removed signal with a 512-D EKYT embedding against the updated setting using a 25 Hz identity-removed signal with a compact 128-D embedding. As shown in Table~\ref{tab:ablation}, the updated configuration improves real-synthetic embedding similarity from $0.910 \pm 0.015$ to $0.956 \pm 0.016$, indicating better preservation of subject-relevant oculomotor structure. Because resampling frequency and embedding dimensionality are changed jointly, this result should be interpreted as a configuration-level comparison rather than an isolation of each individual component.

\begin{table}[h]
\centering
\caption{Ablation comparison of conditioning configurations. Cosine similarity is computed between EKYT embeddings of real and synthetic gaze signals.}
\label{tab:ablation}
\begin{adjustbox}{width=\linewidth}
{\small
\renewcommand{\arraystretch}{1}
\begin{tabular}{ccc}
\hline
Resampling Frequency & Embedding Dimension & Cosine Similarity \\
\hline
20 Hz & 512-D & $0.910 \pm 0.015$ \\
25 Hz & 128-D & $0.956 \pm 0.016$ \\
\hline
\end{tabular}
}
\end{adjustbox}
\end{table}

\section{Discussion}
Our goal was to assess whether high-fidelity diffusion-generated synthetic eye-movement signals preserve privacy-relevant internal states that are known to be decodable from real eye movements. For the original GazeBase recordings, the baseline \cite{qian2025we} reported recurring associations between subjective reports and gaze features such as saccade rate, peak acceleration, fixation rate, and median fixation-drift displacement, especially in pooled analysis and round 2. In contrast, Figures~\ref{fig:hss}(b), \ref{fig:ran}(b), and \ref{fig:tex}(b) show that synthetic gaze exhibits weaker and less stable feature-rating correlations across tasks, with many real-data associations falling below significance or moving closer to zero and only localized residual effects remaining, particularly for TEX in round 2. A reasonable explanation is generative model design, in which the conditioning signal preserves coarse temporal structure and the compact 128-D embedding guides identity and overall signal characteristics. This combination appears to retain person-specific and task-level oculomotor structure, consistent with the high cosine similarities in Table~\ref{tab:cosine_similarity}, while reducing finer-grained trial-level variations that may carry subjective-state information.

From a privacy perspective, this attenuation is desirable because weaker feature-rating associations suggest less exposure of measured subjective-state signatures in the generated signals. This is especially important for gaze biometrics, where eye movements can provide useful identity-related features while also revealing private attributes. Our results, therefore, indicate a privacy-utility trade-off: synthetic gaze remains similar to the original recordings and preserves subject-relevant structure, but its measured associations with subjective states are substantially weakened. However, these findings should be interpreted as evidence of state-signature attenuation under the evaluated protocol, not as complete removal of state information.

From a biometrics perspective, the results are encouraging, suggesting that synthetic gaze can retain utility-relevant subject structure while reducing measured state-related leakage. At the same time, this attenuation limits applications that depend on state decoding. Tasks such as gaze-based fatigue monitoring, cognitive load estimation, or perceived difficulty prediction rely on the same feature-rating associations that are weakened in the synthetic data. Thus, under the current configuration, synthetic gaze may be appropriate for signal-level analysis and biometric-oriented evaluation, but less suitable for downstream tasks that require accurate recovery of transient internal states.

More broadly, the results indicate that the generator does not preserve all components of gaze sequences equally. Identity-related and task-related structure appears to survive the synthesis process more strongly than subjective state-related variability. This distinction is important because high signal realism and high embedding similarity do not necessarily imply preservation of all privacy-relevant latent features. Future work could explicitly control this trade-off using auxiliary losses, disentangled latent representations, or adversarial objectives that separately model identity, task, and state factors. Such approaches may help synthetic gaze remain useful for biometric benchmarking while further reducing the risk of exposing sensitive user states.

Several limitations remain. First, because GazeBase has high-frequency signals, the findings may not generalize to lower-frequency sensors, mobile eye trackers, VR/AR headsets, or unconstrained gaze-interaction settings. Second, the coarse, floor-affected 1-7 Likert ratings may limit the sensitivity of both correlation- and classifier-based state-leakage estimates. Third, the study evaluates attenuation of measured state signatures, not formal privacy guarantees. Accordingly, future work should test stronger state-inference models across broader datasets, sensors, and tasks.

\section{Conclusions}
In this work, we investigated whether diffusion-generated synthetic gaze preserves associations between eye-movement features and subjective reports of internal states, such as fatigue and task difficulty. Using GazeBase as our testbed, we generated synthetic eye movement sequences with an updated DiffEyeSyn framework and computed Spearman correlations between 58 interpretable gaze features and three subjective ratings across horizontal saccades, random oblique saccades, and reading. Overall, synthetic gaze exhibited weaker and less reliable feature-rating associations than real gaze, with only limited residual effects in specific rounds and tasks. The classifier-based validation further showed weak and inconsistent recoverability of binary state under a subject-disjoint protocol. These findings support describing the proposed approach as privacy-aware synthetic gaze generation that attenuates measured state signatures while preserving task-relevant oculomotor structure, rather than as a method that provides formal or complete privacy protection.

{\small
\bibliographystyle{ieee}
\bibliography{egbib}

@String{Computing = "Computing" }

@String{Computer = "{IEEE} Computer" }

@String{Springer = "Springer-Verlag" }

@article{jin2024eye,
  title={Eye-tracking in ar/vr: A technological review and future directions},
  author={Jin, Xin and Chai, Suyu and Tang, Jie and Zhou, Xianda and Wang, Kai},
  journal={IEEE Open Journal on Immersive Displays},
  year={2024},
  publisher={IEEE}
}

@article{griffith2021gazebase,
  title={GazeBase, a large-scale, multi-stimulus, longitudinal eye movement dataset},
  author={Griffith, Henry and Lohr, Dillon and Abdulin, Evgeny and Komogortsev, Oleg},
  journal={Scientific Data},
  volume={8},
  number={1},
  pages={184},
  year={2021},
  publisher={Nature Publishing Group UK London}
}

@article{haller2022eye,
  title={Eye-tracking based classification of Mandarin Chinese readers with and without dyslexia using neural sequence models},
  author={Haller, Patrick and S{\"a}uberli, Andreas and Kiener, Sarah Elisabeth and Pan, Jinger and Yan, Ming and J{\"a}ger, Lena},
  journal={arXiv preprint arXiv:2210.09819},
  year={2022}
}

@article{thanarajan2023eye,
  title={Eye-Tracking Based Autism Spectrum Disorder Diagnosis Using Chaotic Butterfly Optimization with Deep Learning Model.},
  author={Thanarajan, Tamilvizhi and Alotaibi, Youseef and Rajendran, Surendran and Nagappan, Krishnaraj},
  journal={Computers, Materials \& Continua},
  volume={76},
  number={2},
  year={2023}
}

@article{lohr2022ekyt,
  title={Eye know you too: Toward viable end-to-end eye movement biometrics for user authentication},
  author={Lohr, Dillon and Komogortsev, Oleg V},
  journal={IEEE Transactions on Information Forensics and Security},
  volume={17},
  pages={3151--3164},
  year={2022},
  publisher={IEEE}
}

@inproceedings{jiao2023supreyes,
  title={SUPREYES: SUPer Resolutin for EYES Using Implicit Neural Representation Learning},
  author={Jiao, Chuhan and Hu, Zhiming and B{\^a}ce, Mihai and Bulling, Andreas},
  booktitle={Proceedings of the 36th Annual ACM Symposium on User Interface Software and Technology},
  pages={1--13},
  year={2023}
}

@inproceedings{steil2019privacy,
  title={Privacy-aware eye tracking using differential privacy},
  author={Steil, Julian and Hagestedt, Inken and Huang, Michael Xuelin and Bulling, Andreas},
  booktitle={Proceedings of the 11th ACM Symposium on Eye Tracking Research \& Applications},
  pages={1--9},
  year={2019}
}

@inproceedings{duchowski2015modeling,
  title={Modeling physiologically plausible eye rotations},
  author={Duchowski, Andrew T and J{\"o}rg, Sophie},
  booktitle={Proceedings of computer graphics international},
  year={2015}
}

@inproceedings{assens2018pathgan,
  title={PathGAN: Visual scanpath prediction with generative adversarial networks},
  author={Assens, Marc and Giro-i-Nieto, Xavier and McGuinness, Kevin and O'Connor, Noel E},
  booktitle={Proceedings of the European Conference on Computer Vision (ECCV) Workshops},
  pages={0--0},
  year={2018}
}

@inproceedings{simon2016automatic,
  title={Automatic scanpath generation with deep recurrent neural networks},
  author={Simon, Daniel and Sridharan, Srinivas and Sah, Shagan and Ptucha, Raymond and Kanan, Chris and Bailey, Reynold},
  booktitle={Proceedings of the ACM symposium on applied perception},
  pages={130--130},
  year={2016}
}

@inproceedings{duchowski2016eye,
  title={Eye movement synthesis},
  author={Duchowski, Andrew T and J{\"o}rg, Sophie and Allen, Tyler N and Giannopoulos, Ioannis and Krejtz, Krzysztof},
  booktitle={Proceedings of the ninth biennial ACM symposium on eye tracking research \& applications},
  pages={147--154},
  year={2016}
}

@article{yeo2012eyecatch,
  title={Eyecatch: Simulating visuomotor coordination for object interception},
  author={Yeo, Sang Hoon and Lesmana, Martin and Neog, Debanga R and Pai, Dinesh K},
  journal={ACM Transactions on Graphics (TOG)},
  volume={31},
  number={4},
  pages={1--10},
  year={2012},
  publisher={ACM New York, NY, USA}
}

@inproceedings{zhang2023adding,
  title={Adding conditional control to text-to-image diffusion models},
  author={Zhang, Lvmin and Rao, Anyi and Agrawala, Maneesh},
  booktitle={Proceedings of the IEEE/CVF international conference on computer vision},
  pages={3836--3847},
  year={2023}
}

@inproceedings{lee2002eyes,
  title={Eyes alive},
  author={Lee, Sooha Park and Badler, Jeremy B and Badler, Norman I},
  booktitle={Proceedings of the 29th annual conference on Computer graphics and interactive techniques},
  pages={637--644},
  year={2002}
}

@article{lohr2025gaze,
  title={Gaze Authentication: Factors Influencing Authentication Performance},
  author={Lohr, Dillon and Proulx, Michael J and Raju, Mehedi Hasan and Komogortsev, Oleg V},
  journal={arXiv preprint arXiv:2509.10969},
  year={2025}
}

@article{goodfellow2020generative,
  title={Generative adversarial networks},
  author={Goodfellow, Ian and Pouget-Abadie, Jean and Mirza, Mehdi and Xu, Bing and Warde-Farley, David and Ozair, Sherjil and Courville, Aaron and Bengio, Yoshua},
  journal={Communications of the ACM},
  volume={63},
  number={11},
  pages={139--144},
  year={2020},
  publisher={ACM New York, NY, USA}
}

@article{jiao2025diffgaze,
  title={DiffGaze: A Diffusion Model for Modelling Fine-grained Human Gaze Behaviour on 360° Images},
  author={Jiao, Chuhan and Wang, Yao and Zhang, Guanhua and B{\^a}ce, Mihai and Hu, Zhiming and Bulling, Andreas},
  journal={ACM Transactions on Interactive Intelligent Systems},
  year={2025},
  publisher={ACM New York, NY}
}

@inproceedings{prasse2023sp,
  title={Sp-eyegan: Generating synthetic eye movement data with generative adversarial networks},
  author={Prasse, Paul and Reich, David Robert and Makowski, Silvia and Ahn, Seoyoung and Scheffer, Tobias and J{\"a}ger, Lena A},
  booktitle={Proceedings of the 2023 symposium on eye tracking research and applications},
  pages={1--9},
  year={2023}
}

@article{ho2020denoising,
  title={Denoising diffusion probabilistic models},
  author={Ho, Jonathan and Jain, Ajay and Abbeel, Pieter},
  journal={Advances in neural information processing systems},
  volume={33},
  pages={6840--6851},
  year={2020}
}

@article{loshchilov2017decoupled,
  title={Decoupled weight decay regularization},
  author={Loshchilov, Ilya and Hutter, Frank},
  journal={arXiv preprint arXiv:1711.05101},
  year={2017}
}

@article{hasan2026diffusion,
  title={Diffusion Versus GAN for Subject-Specific Gaze Synthesis},
  author={Hasan, Kamrul and Komogortsev, Oleg V},
  journal={IEEE pulse},
  volume={17},
  number={1},
  pages={60--62},
  year={2026},
  publisher={IEEE}
}

@inproceedings{jiao2024diffeyesyn,
  title={DiffEyeSyn: User-specific Subtle Eye Movement Synthesis Using Diffusion Models},
  author={Jiao, Chuhan and Zhang, Guanhua and Cho, Yeonjoo and Hu, Zhiming and Bulling, Andreas},
  booktitle={2026 IEEE 20th International Conference on Automatic Face and Gesture Recognition (FG)},
  pages={1--10},
  year={2026},
  organization={IEEE}
}

@article{raju2021determining,
  title={Determining which sine wave frequencies correspond to signal and which correspond to noise in eye-tracking time-series},
  author={Raju, Mehedi H and Friedman, Lee and Bouman, Troy M and Komogortsev, Oleg V},
  journal={Journal of Eye Movement Research},
  volume={14},
  number={3},
  pages={16},
  year={2021},
  publisher={Bern Open Publishing}
}

@inproceedings{aziz2024evaluation,
  title={Evaluation of eye tracking signal quality for virtual reality applications: A case study in the meta quest pro},
  author={Aziz, Samantha and Lohr, Dillon J and Friedman, Lee and Komogortsev, Oleg},
  booktitle={Proceedings of the 2024 Symposium on Eye Tracking Research and Applications},
  pages={1--8},
  year={2024}
}

@article{hayhoe2005eye,
  title={Eye movements in natural behavior},
  author={Hayhoe, Mary and Ballard, Dana},
  journal={Trends in cognitive sciences},
  volume={9},
  number={4},
  pages={188--194},
  year={2005},
  publisher={Elsevier}
}

@inproceedings{sun2017application,
  title={The application of eye tracking in education},
  author={Sun, Yuyang and Li, Qingzhong and Zhang, Honggen and Zou, Jiancheng},
  booktitle={International Conference on Intelligent Information Hiding and Multimedia Signal Processing},
  pages={27--33},
  year={2017},
  organization={Springer}
}

@article{zar2005spearman,
  title={Spearman rank correlation},
  author={Zar, Jerrold H},
  journal={Encyclopedia of biostatistics},
  volume={7},
  year={2005},
  publisher={Wiley Online Library}
}

@book{maltoni2009handbook,
  title={Handbook of fingerprint recognition},
  author={Maltoni, Davide and Maio, Dario and Jain, Anil K and Prabhakar, Salil},
  year={2009},
  publisher={Springer}
}

@article{zhao2003face,
  title={Face recognition: A literature survey},
  author={Zhao, Wenyi and Chellappa, Rama and Phillips, P Jonathon and Rosenfeld, Azriel},
  journal={ACM computing surveys (CSUR)},
  volume={35},
  number={4},
  pages={399--458},
  year={2003},
  publisher={ACM New York, NY, USA}
}

@article{uddin2024horizontal,
  title={Horizontal and vertical part-wise feature extraction for cross-view gait recognition},
  author={Uddin, Md Zasim and Hasan, Kamrul and Ahad, Md Atiqur Rahman and Alnajjar, Fady},
  journal={IEEE Access},
  volume={12},
  pages={185511--185527},
  year={2024},
  publisher={IEEE}
}

@article{qian2025we,
  title={Why do we need high-fidelity synthetic eye movement data and how should they look like?},
  author={Qian, C Stella and Aziz, Samantha and Hasan, Kamrul and Komogortsev, Oleg V},
  journal={bioRxiv},
  pages={2025--12},
  year={2025},
  publisher={Cold Spring Harbor Laboratory}
}

@article{hasan2025quantitative,
  title={Quantitative and Qualitative Comparison of Generative Models for Subject-Specific Gaze Synthesis: Diffusion vs GAN},
  author={Hasan, Kamrul and Katrychuk, Dmytro and Raju, Mehedi Hasan and Komogortsev, Oleg V},
  journal={arXiv preprint arXiv:2511.09867},
  year={2025}
}

@inproceedings{aziz2023assessing,
  title={Assessing the privacy risk of cross-platform identity linkage using eye movement biometrics},
  author={Aziz, Samantha and Komogortsev, Oleg},
  booktitle={2023 IEEE International Joint Conference on Biometrics (IJCB)},
  pages={1--9},
  year={2023},
  organization={IEEE}
}

@inproceedings{david2022your,
  title={For your eyes only: Privacy-preserving eye-tracking datasets},
  author={David-John, Brendan and Butler, Kevin and Jain, Eakta},
  booktitle={2022 symposium on eye tracking research and applications},
  pages={1--6},
  year={2022}
}

@article{bozkir2021differential,
  title={Differential privacy for eye tracking with temporal correlations},
  author={Bozkir, Efe and G{\"u}nl{\"u}, Onur and Fuhl, Wolfgang and Schaefer, Rafael F and Kasneci, Enkelejda},
  journal={Plos one},
  volume={16},
  number={8},
  pages={e0255979},
  year={2021},
  publisher={Public Library of Science San Francisco, CA USA}
}

@inproceedings{liu2019differential,
  title={Differential privacy for eye-tracking data},
  author={Liu, Ao and Xia, Lirong and Duchowski, Andrew and Bailey, Reynold and Holmqvist, Kenneth and Jain, Eakta},
  booktitle={Proceedings of the 11th ACM Symposium on Eye Tracking Research \& Applications},
  pages={1--10},
  year={2019}
}

@article{aziz2025privacy,
  title={Privacy Enhancement for Gaze Data Using a Noise-Infused Autoencoder},
  author={Aziz, Samantha and Komogortsev, Oleg},
  journal={arXiv preprint arXiv:2508.10918},
  year={2025}
}

@inproceedings{kroger2020does,
  title={What does your gaze reveal about you? On the privacy implications of eye tracking},
  author={Kr{\"o}ger, Jacob Leon and Lutz, Otto Hans-Martin and M{\"u}ller, Florian},
  booktitle={IFIP International Summer School on Privacy and Identity Management},
  pages={226--241},
  year={2020},
  organization={Springer}
}

@article{rigas2018study,
  title={Study of an extensive set of eye movement features: Extraction methods and statistical analysis},
  author={Rigas, Ioannis and Friedman, Lee and Komogortsev, Oleg},
  journal={Journal of Eye Movement Research},
  volume={11},
  number={1},
  pages={10--16910},
  year={2018}
}

@article{friedman_eye_2022,
	title = {Eye Movement Classification Dataset},
	url = {https://hdl.handle.net/10877/16339},
	author = {Friedman, Lee},
	urldate = {2025-11-07},
	date = {2022-11},
	langid = {english},
}

@article{friedman_novel_2018,
	title = {A novel evaluation of two related and two independent algorithms for eye movement classification during reading},
	volume = {50},
	issn = {1554-3528},
	url = {https://doi.org/10.3758/s13428-018-1050-7},
	doi = {10.3758/s13428-018-1050-7},
	pages = {1374--1397},
	number = {4},
	journaltitle = {Behavior Research Methods},
	shortjournal = {Behav Res},
	author = {Friedman, Lee and Rigas, Ioannis and Abdulin, Evgeny and Komogortsev, Oleg V.},
	urldate = {2025-11-19},
	date = {2018-08-01},
	langid = {english},
}

@article{liu2021effectiveness,
  title={The effectiveness of eye tracking in the diagnosis of cognitive disorders: A systematic review and meta-analysis},
  author={Liu, Zicai and Yang, Zhen and Gu, Yueming and Liu, Huiyu and Wang, Pu},
  journal={PloS one},
  volume={16},
  number={7},
  pages={e0254059},
  year={2021},
  publisher={Public Library of Science San Francisco, CA USA}
}

@article{raju2025real,
  title={Real-Time Lightweight Gaze Privacy-Preservation Techniques Validated via Offline Gaze-Based Interaction Simulation},
  author={Raju, Mehedi Hasan and Komogortsev, Oleg V},
  journal={arXiv preprint arXiv:2511.09846},
  year={2025}
}
}

\end{document}